\documentclass[prl,showpacs,floatfix,preprint]{revtex4}

\usepackage{graphicx} 

\def\bto{Bi$_2$Ti$_2$O$_7$}
\def\bnio{Bi$_2$NbInO$_7$}
\def\yto{Y$_2$Ti$_2$O$_7$}
\def\pto{PbTiO$_3$}
\def\btofe{Bi$_4$Ti$_3$O$_{12}$}

\begin{document} 

\title{Large Low Temperature Specific Heat in Pyrochlore Bi$_2$Ti$_2$O$_7$} 

\author{Brent\,C. Melot,$^1$, Ronald\,Tackett$^2$, Jim\,O'Brien,$^3$, 
Andrew\,L.\,Hector$^4$, Gavin\,Lawes$^2$, Ram\,Seshadri$^1$, 
Arthur\,P.\,Ramirez$^5$}

\affiliation{$^1$Materials Department, 
University of California Santa Barbara, Santa Barbara, CA 93106\\
$^2$Department of Physics and Astronomy, Wayne State University, 
Detroit, MI 48201\\
$^3$Quantum Design, 6325 Lusk Boulevard, San Diego, CA 92121\\
$^4$School of Chemistry, University of Southampton, Highfield, 
Southampton SO17 1BJ, UK\\
$^5$LGS, 15 Vreeland Road, Florham Park, New Jersey NJ 07932\\
            }

\date{\today} 

\begin{abstract}
Both amorphous and crystalline materials frequently exhibit low temperature 
specific heats in excess of what is predicted using the Debye model. 
The signature of this excess specific heat is a peak observed in 
$C/T^3$ \textit{versus} $T$. To understand the curious absence 
of long-range ordering of local distortions in the crystal structure of 
pyrochlore \bto, we have measured the specific heat of crystalline \bto\, and
related compounds. We find that the peak in $C/T^3$ \textit{versus} $T$
in \bto\, falls at a substantially lower temperature than other 
similar compounds, consistent with the presence of disorder.
This thermodynamic evidence for disorder in crystalline \bto\, 
is consistent with quenched configurational disorder 
among Bi lone pairs produced by geometrical frustration, 
which could represent a possible realization of ``charge ice''.
\end{abstract} 

\maketitle 

It is well known that at relatively low temperatures, typically in the range 
of 2\,K to 30\,K, amorphous systems exhibit a larger specific heat than the 
simple prediction of the Debye model. This excess specific heat manifests 
itself as a peak in $C/T^3$ \textit{versus} $T$, and is generally attributed to 
local low energy vibrational modes not accounted for in the Debye model. 
These low energy modes are observed in Raman 
spectra\cite{Winterling_PRB75} and from inelastic neutron 
scattering,\cite{Buchenau_PRB86} where they give rise, in glasses, to the 
so-called ``boson peak''.\cite{Zeller_PRB71,Phillips_JLTP72} The mechanisms 
responsible for these low energy modes have been discussed in terms of 
localized vibrations (``floppy modes''), domain wall motions of the glassy
mosaic structure, and transverse phonon 
modes.\cite{Dove_PRL97,Lubchenko_PNAS03,Shintani_NatureMater08} 

A number of crystalline materials also exhibit a low-temperature peak in 
$C/T^3$. This peak can be attributed to van Hove singularities where the 
vibrational density of states (VDOS) crosses the Debye density of 
states, leading to a flattening of phonon dispersion curve.\cite{Safarik_PRL06}
 This is nearly equivalent to stating that the 
local modes are responsible; in materials such as SiO$_2$, the
nature of the vibrations responsible for the peak in crystalline and amorphous
samples are similar.\cite{Dove_PRL97} 
 As low temperature specific heat measurements can provide
 evidence for disorder in insulating crystalline 
compounds, we have used thermodynamic measurements to 
investigate displacive disorder in \bto\, and related systems. 

\begin{figure}
\centering\includegraphics[width=8.5cm]{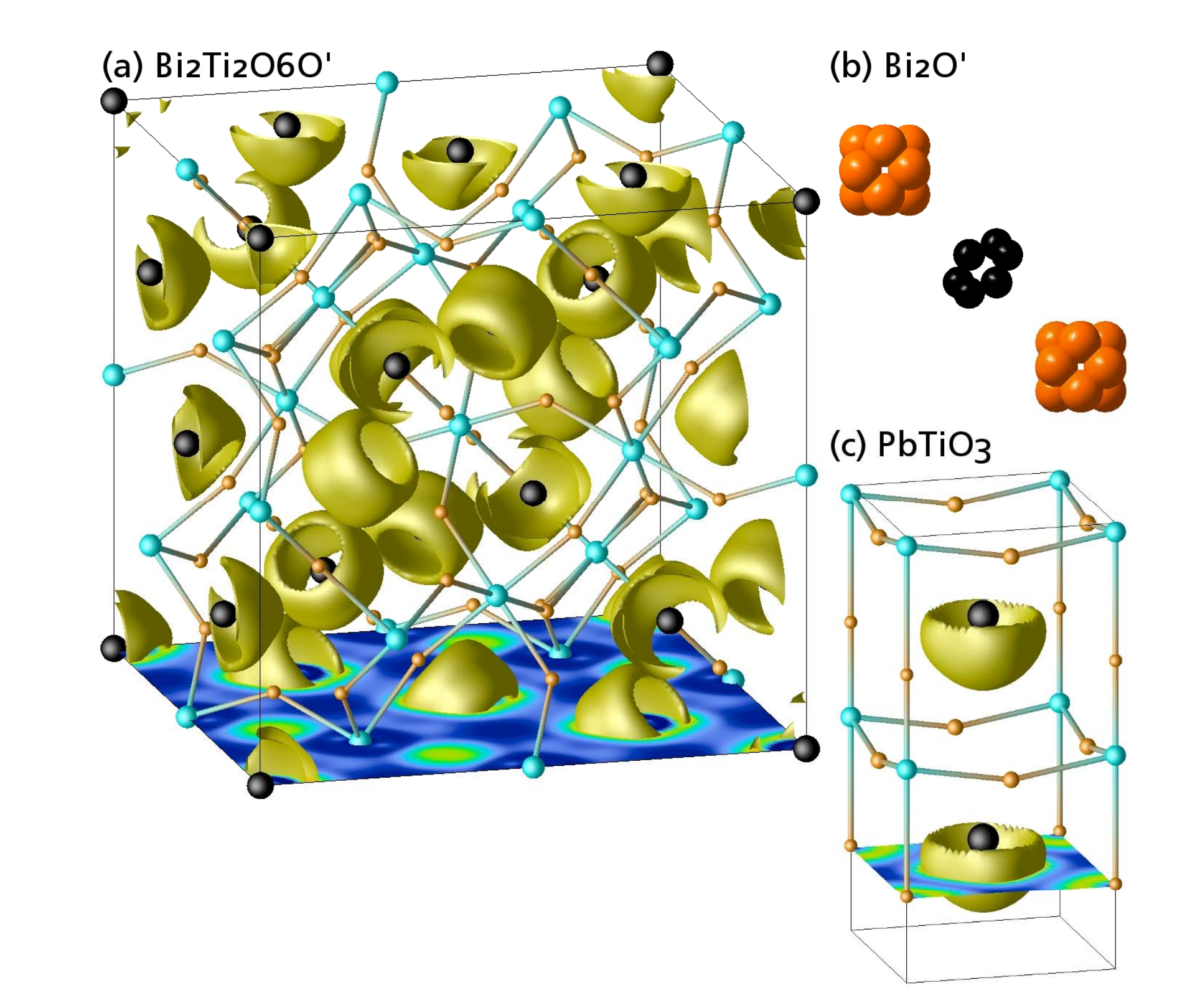}\\
\caption{(a) Structure of ideal Bi$_2$Ti$_2$O$_6$O$^\prime$. The 6$s^2$ lone 
pair around Bi$^{3+}$ atoms are visualized using the valence electron 
localization function (ELF) with isosurfaces of value 0.625 following methods 
presented in reference \onlinecite{Seshadri_SSC06}. (b) Split atom positions
of Bi and O$^\prime$ (50\% isotropic thermal ellipsoids displayed) 
as obtained from Rietveld refinement of powder neutron 
diffraction data indicative of local displacive disorder, following 
reference \onlinecite{Melot_MRB06}. (c) Crystal structure of tetragonal 
PbTiO$_3$ (two unit cells displayed), using the ELF to locate lone pairs 
around Pb$^{2+}$. Note the difference in the nature of the lone pairs
in (a) and (c). Black spheres are Bi or Pb, cyan spheres are Ti, and orange 
spheres are O.}
\label{fig:structures}
\end{figure}

In this contribution, we examine the  excess low temperature specific heat in 
a series of structurally- and compositionally-related crystalline compounds of 
interest as polar and dielectric materials. The pyrochlore
compound Bi$_2$Ti$_2$O$_6$O$^\prime$ is compositionally and electronically
related to the perovskite ferroelectric PbTiO$_3$, in that the A site of the 
structure is occupied by a main group ion with the 6$s^2,6p^0$ configuration, 
and the B site is occupied by $d^0$ Ti$^{4+}$ (structures displayed in 
FIG.\,\ref{fig:structures}). Both the A and the B site ions in these compounds
are therefore susceptible to off-centering.  This so-called stereochemical
activity, and this is what drives the phase transition to a coherent, polar,
tetragonal ground state in PbTiO$_3$. The related pyrochlore compound
Bi$_2$Ti$_2$O$_6$O$^\prime$ (more simply referred to as \bto) shows no such
phase transition and even at 2\,K, the crystal structure is 
cubic.\cite{Hector_JSSC04} However, in a manner that is common across many
pyrochlore crystal structures with Bi$^{3+}$ on the A site, Rietveld
refinement of neutron diffraction data indicates that Bi is locally
off-centered.  This evidence for local distortion, in conjuction
with the absence of ferroelectricity, suggests that that the Bi ions are displaced randomly into one of
 several distinct, but symmetric,
sites about the equilibrium position as indicated in 
FIG.\,\ref{fig:structures}(b).\cite{Avdeev_JSSC02,Levin_JSSC02,Hector_JSSC04,Zhou_JSSC04,Melot_MRB06,Rodriguez_InorgChem08}

The goal of this work is to explore whether there is any thermodynamic
evidence for displacive disorder in \bto\, driven by geometric 
frustration on the pyrochlore lattice.\cite{Seshadri_SSC06,McQueen_JPCM08} 
Frustration of ferromagnetic Ising spins in compounds with the pyrochlore 
crystal structure are well known,\cite{Ramirez_Nature99,Bramwell_Science02} 
with analogies having being drawn with the problem of hydrogen positions
in the crystal structure of cubic ice $I_c$, with its associated residual 
entropy.\cite{Pauling_JACS35} A similar 
analogy between ice and putative \textit{polar} pyrochlores was suggested 
in reference \onlinecite{Seshadri_SSC06}; that \bto\, may be a manifestation of 
``charge ice''. We emphasize that this disorder is expected to be driven by
purely geometrical considerations and develop on well-ordered crystalline
lattices, so this ``charge ice'' structure should be distinguished from the
``charge glass'' state observed in structurally disordered La$_2$Cu$_{1-x}$Li$_x$O$_4$
and La$_{2-x}$Sr$_x$NiO$_4$.\cite{Park_PRL05}

In order investigate the role of disorder on the low temperature
heat capacity in crystalline systems we have measured the specific heat 
of \bto,  two isostructural compounds \yto\, and \bnio, as well as the
ferroelectrics \pto\, and \btofe. Coherent lone pair displacements in the
 latter two systems produce a ferroelectric ground state, which suggests that residual displacive 
disorder should be minimal. The compound \yto\, unlike the 
four others, has no lone pairs on the A site and can be expected to display the 
least disorder. All the samples are white to light yellow and highly 
electrically insulating in powder form.

The polycrystalline samples studied here, with the exception of \bto,
were prepared by ceramic routes and phase purity was verified using powder
X-ray diffraction. Perovskite PbTiO$_3$ was prepared from PbO and TiO$_2$, with
a 5\% stoichiometric excess of PbO to balance losses due to volatilization.
Pyrochlore Y$_2$Ti$_2$O$_7$ was prepared from Y$_2$O$_3$ and TiO$_2$.
The Aurivilius phase Bi$_4$Ti$_3$O$_{12}$ was obtained from appropriate
stoichiometric starting ratios of Bi$_2$O$_3$ and TiO$_2$.
Bi$_2$InNbO$_7$ was prepared  by reacting appropriate stoichiometric
amounts of Bi$_2$O$_3$, Nb$_2$O$_5$, and In$_2$O$_3$, following the method
reported by Zhou \textit{et al.}\cite{Zhou_JSSC04} The \bto\/ sample used in 
this study was prepared by low temperature routes and previously subject to 
neutron structure determination.\cite{Hector_JSSC04} All specific heat 
measurements were carried out on Quantum Design Physical Properties Measurement
Systems (PPMS). We used different techniques to extract the low temperature 
specific heat of these compounds. The \bto, \pto, and \btofe\, powders were 
pressed into dense pellets, mounted to a calorimeter with thermal grease.
The \bnio\, powder was cold sintered with Ag powder, with the specific heat contribution
from the silver measured separately and subtracted.  The specific heat of 
\yto\, was measured both by mixing the loose powder with a small amount of 
thermal grease then compacted between aluminum plates
with weighing paper and also by pressing it in a 1:1 ratio with silver. Both 
techniques yielded quantitively similar results, so we present only the data 
from \yto\, mixed with thermal grease. The \bto\, and \yto\, samples
were separately measured down to 0.5\,K using a PPMS system equipped with a 
$^3$He refrigerator, with the relaxation measured over two time-constants.

\begin{figure}
\centering\includegraphics[width=8.5cm]{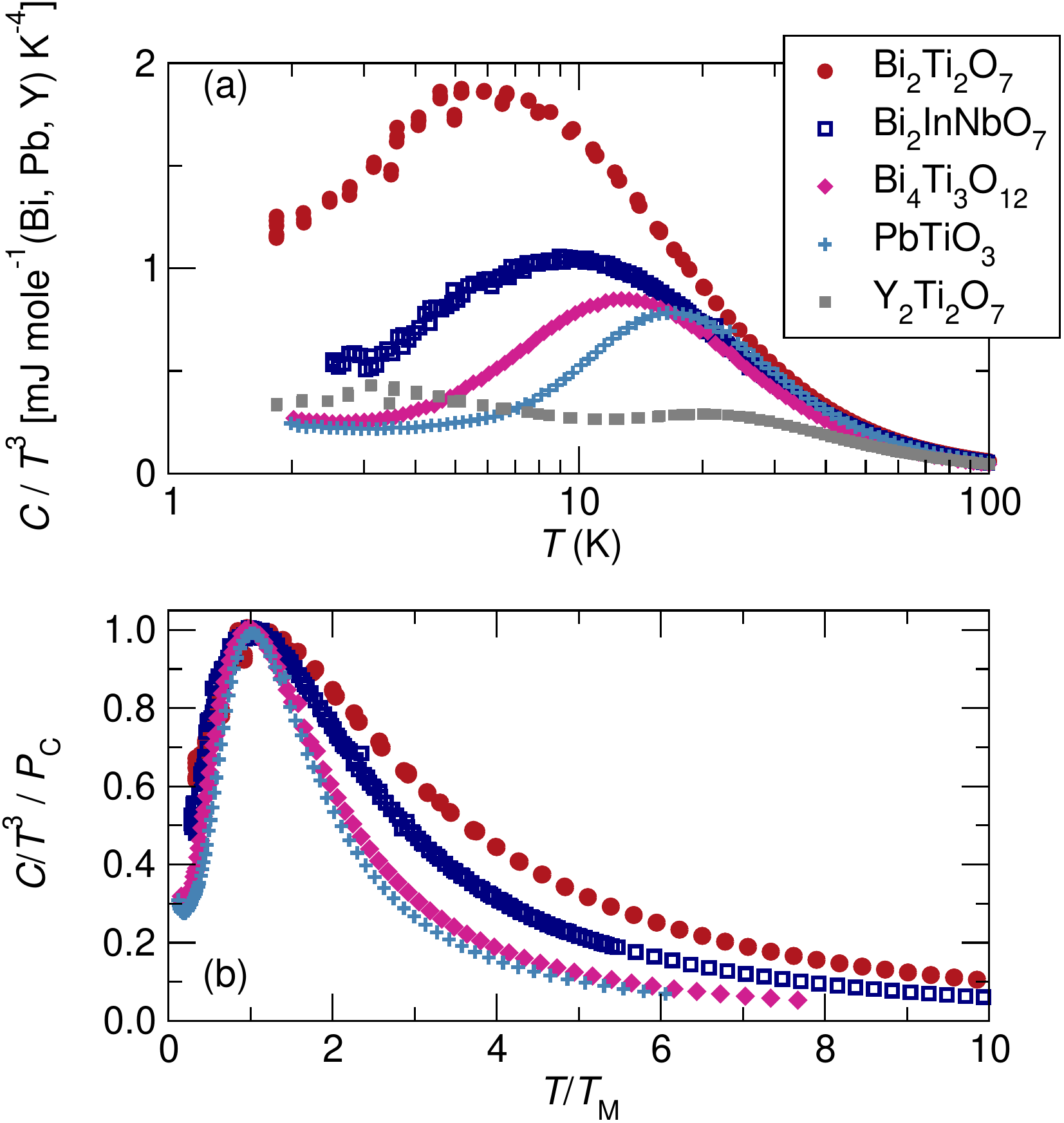}\\
\caption{(color on-line) (a) Temperature dependence of the low temperature specific heat 
(plotted as $C/T^3$ \textit{versus} $T$) for the five samples studied in this 
contribution. The values are per mole of the A-site atoms in the 
different formulae. For all samples, the values of $C/T^3$ in excess of 
about 0.5\,mJ\,mole$^{-1}$\,K$^{-4}$ are contributions not accounted for by 
the Debye model. (b) Scaled plots: $C/T^3/P_{\rm C}$ as a function of the 
scaled  temperature $T/T_{\rm M}$.}
\label{fig:heatcapacity}
\end{figure}

The specific heat capacities for all five samples are
shown in FIG.\,\ref{fig:heatcapacity}(a). In order to more clearly emphasize 
the low energy features in the system, giving rise to specific heat 
in excess of the Debye contribution, we have plotted the
specific heat as $C/T^3$ \textit{versus} $T$.  All samples show a
maximum when plotted in this manner, with the amplitude of the
peak increasing monotonically with decreasing peak temperature. 
This peak indicates that the specific heat of these crystalline
samples exceeds the Debye C(T) at low temperatures, and
are consistent with previous measurements on crystalline \pto,
which has been shown to exhibit a peak at $T_{\rm M}$ = 15\,K.\cite{PTO_HC}

As \yto\, and \bto, have the same crystal structure, one would expect that both compounds
should exhibit 
similar vibrational DOS. Contrary to this expectation, we find that \bto\, has a much 
larger low energy density of states, which is partly due to the larger mass of Bi compared to Y.  This larger mass 
cannot, however, fully explain the appearance of a low-energy peak;  we attribute this excess heat capacity to the 
presence of disorder among Bi lone pairs.
 The lower temperature for the \bto\, peak 
relative to the \yto\, peak is consistent with increasing
disorder in the Bi pyrochlore
arising from lone pair displacements, as discussed in more detail in the following.
  The other pyrochlore compound studied here, which is also expeced to
displacive disorder on the A site, \bnio,  shows a relatively large, 
low-temperature peak in $C/T^3$, although the amplitude of the peak, 
$P_{\rm C}$ is smaller, and the temperature $T_{\rm M}$ at which the peak 
maximum is found is higher that \bto. 

In order to compare the excess specific heat in the
two pyrochlore compounds with other crystalline lone-pair active systems, 
we have also measured the low temperature specific heat of \pto\, and 
\btofe. It is seen in FIG.\,\ref{fig:heatcapacity}(a)
that these compounds also display a distinct peak in $C/T^3$, although the peaks fall at higher energy, and are
smaller in magnitude than the peaks displayed by the two pyrochlore compounds with
displacive disorder.
The distinct difference in the excess specific heat of the lone-pair disordered
compounds, \bto\, and \bnio\, from the lone-pair ordered compounds,
\pto\, and \btofe\, is seen in the scaled $C/T^3$ \textit{vs.} $T$ plots
displayed FIG.\,\ref{fig:heatcapacity}(b). The scaling was performed with 
respect to the individual peaks, $P_C$ at which $C/T^3$ is maximum, and the
temperature $T_M$ where the maximum is found. While materials having both
ordered and disordered distortions exhibit a specific heat peak, the lone-pair 
disordered compounds display a distinctly greater width in the distribution of 
the excess specific heat.  This increase in the scaled full width half maximum 
(FWHM) of the \bto\, and \bnio\, curves indicates that the modes contributing 
to the excess entropy in these materials have a wider distributions of 
energies than in the ordered compounds.  As a reference, the curve
computed for a single mode Einstein oscillator is plotted as a solid line in
 FIG.\,\ref{fig:heatcapacity}(b). The relative width of this excess heat capacity peak, 
 plotted on a semilog scale and measured at 70\% of the maximum value, increases from 0.36 
 for \pto\, to 0.71 for \bnio\, and 1.05 for \bto.  This increase is similar to the broadening 
 of the $C/T^3$ maximum in amorphous SiO$_2$ relative to crystalline SiO$_2$\cite{Liu_EurophysLett96}, 
 and can be associated with increasing disorder of the lone pair electrons in \bnio\, and \bto.

\begin{figure}
\centering\includegraphics[width=8.5cm]{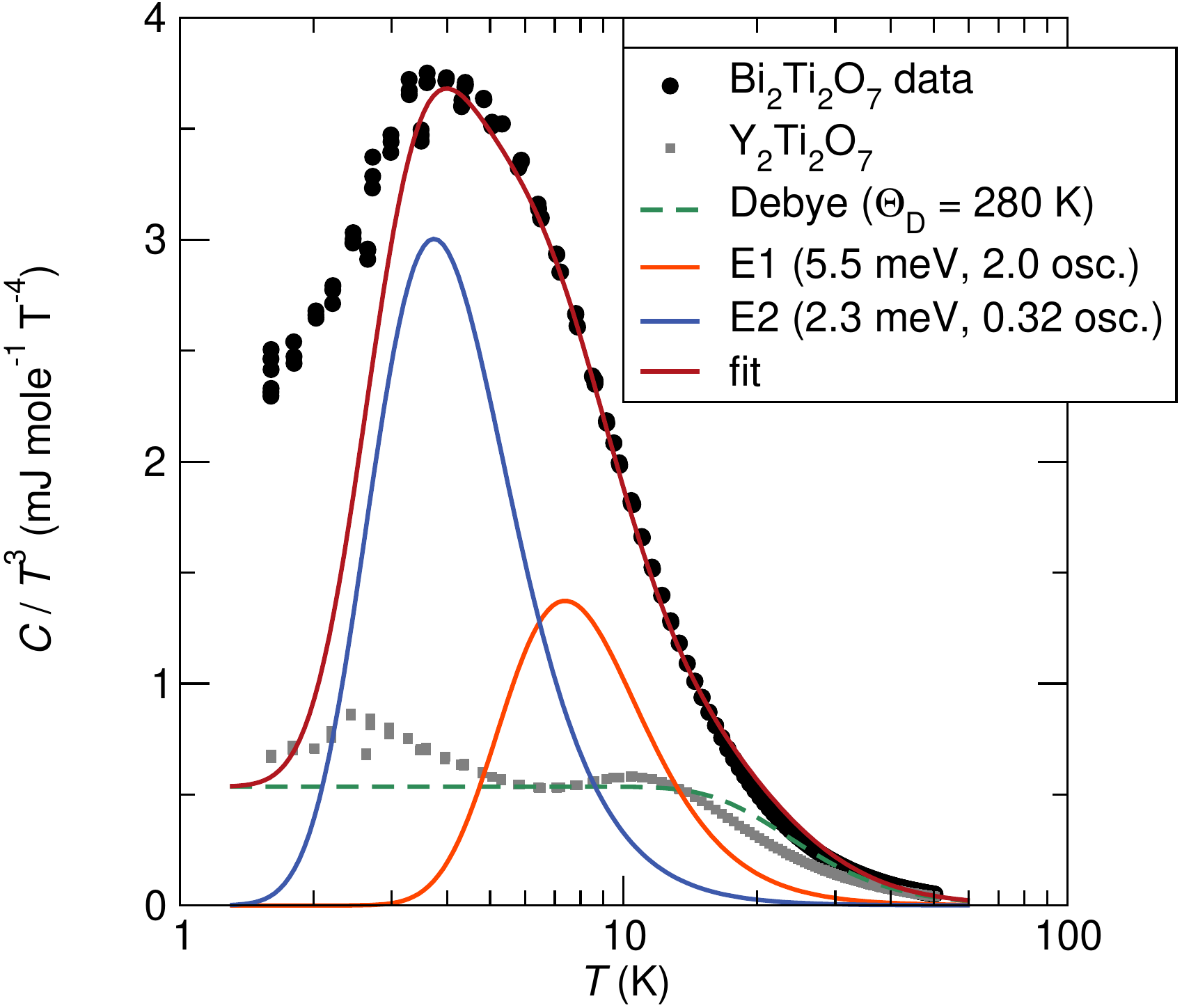}\\
\caption{(color on-line) Fits to the specific heat of \bto\, using a combinations
of Debye and Einstein modes, as described in the text. Data for \yto\/
are seen to be almost completely fit by the single Debye mode
using the same value of $\Theta_D$ used for \bto.}
\label{fig:BTO_fit}
\end{figure}

To quantitatively investigate these peaks, we fit the specific heat in 
FIG.\,\ref{fig:BTO_fit} to the sum of a Debye contribution ($C_D$), 
with two additional low energy Einstein modes ($C_E$). This is equivalent 
to supposing that the VDOS can be modeled by a delta function as it
crosses the Debye DOS, rather than the finite jump expected for
van Hove singularities in 3D crystals. We included two separate
Einstein modes as FIG.\,\ref{fig:heatcapacity}(b) indicates that the 
excess heat capacity cannot be fit by a single mode.  While simplistic, this
model allows us to relate the thermodynamic anomaly to microscopic
properties in the samples.  For all crystalline samples included in this study
(\bto, \yto, \pto, \bnio, and \btofe) we found that a suitable fit could be obtained
using a Debye temperature of $\theta_D=280$ K.  This emphasizes that the
underlying lattice heat capacities of these different samples are similar, so that
the significant differences in the low temperature heat capacities should
be attributed to differences in the lone pair electron behaviour. 

The low temperature $C(T)$ can be modeled as a sum of Debye and Einstein terms with each 
Einstein oscillator giving a contribution: 

\begin{equation}
C_E = pR\frac{(\hbar\omega_0/k_BT)^2e^{\hbar\omega_0/k_BT}}
                         {(e^{\hbar\omega_0/k_BT}-1)^2}
\end{equation}

\noindent to the specific heat. In this expression, $p$ is the spectral 
weight, $R$ the gas constant, and $\hbar\omega_0$ the mode energy. For 
\bto\, the first excitation has an energy of $E_1$ = 2.3\,meV and an 
oscillator strength of $p_1$ = 0.32, while the second excitation falls at 
$E_2$ = 5.5\,meV  and has an oscillator strength of $p_2$ = 2. Similar fits 
were obtained for the other samples. We find that  \bto\, has the lowest 
energy oscillator. The values of the oscillator strengths and energies for the 
different samples are given in Table\,\ref{Table}. We find that 
$\hbar\omega_0$ for the lowest energy mode in these fits depends linearly on 
the temperature of the peak. Motivated by this result, which, within the 
constraints of our simple model, confirms that the peak temperature and 
vibrational mode frequencies are linearly related, we will discuss our data 
solely in terms of the peak temperature. This assumption allows us to avoid 
specific details of the fitting parameters, allowing a more general and robust 
analysis of the results.

\begin{table}
\begin{tabular}{|l | l | l | l | l |}
\hline
\hline
  & $E_1$ (meV) & $p_1$ & $E_2$ (meV) & $p_2$ \\ \hline
 \bto & 2.3 & 0.32 & 5.5 & 2 \\ 
 \btofe & 3.6 & 0.7 & 6.7 & 5 \\
 \bnio & 2.8 & 0.27 & 6 & 1.6 \\ 
 \pto & 5.5 & 0.4 & 8 & 1.5 \\ 
\hline
\hline
\end{tabular}
\caption{Energies and number of oscillators used in fitting Einstein modes
to the specific heats of the diferent compounds.
\label{Table}}
\end{table}

\begin{figure}
\centering\includegraphics[width=8.5cm]{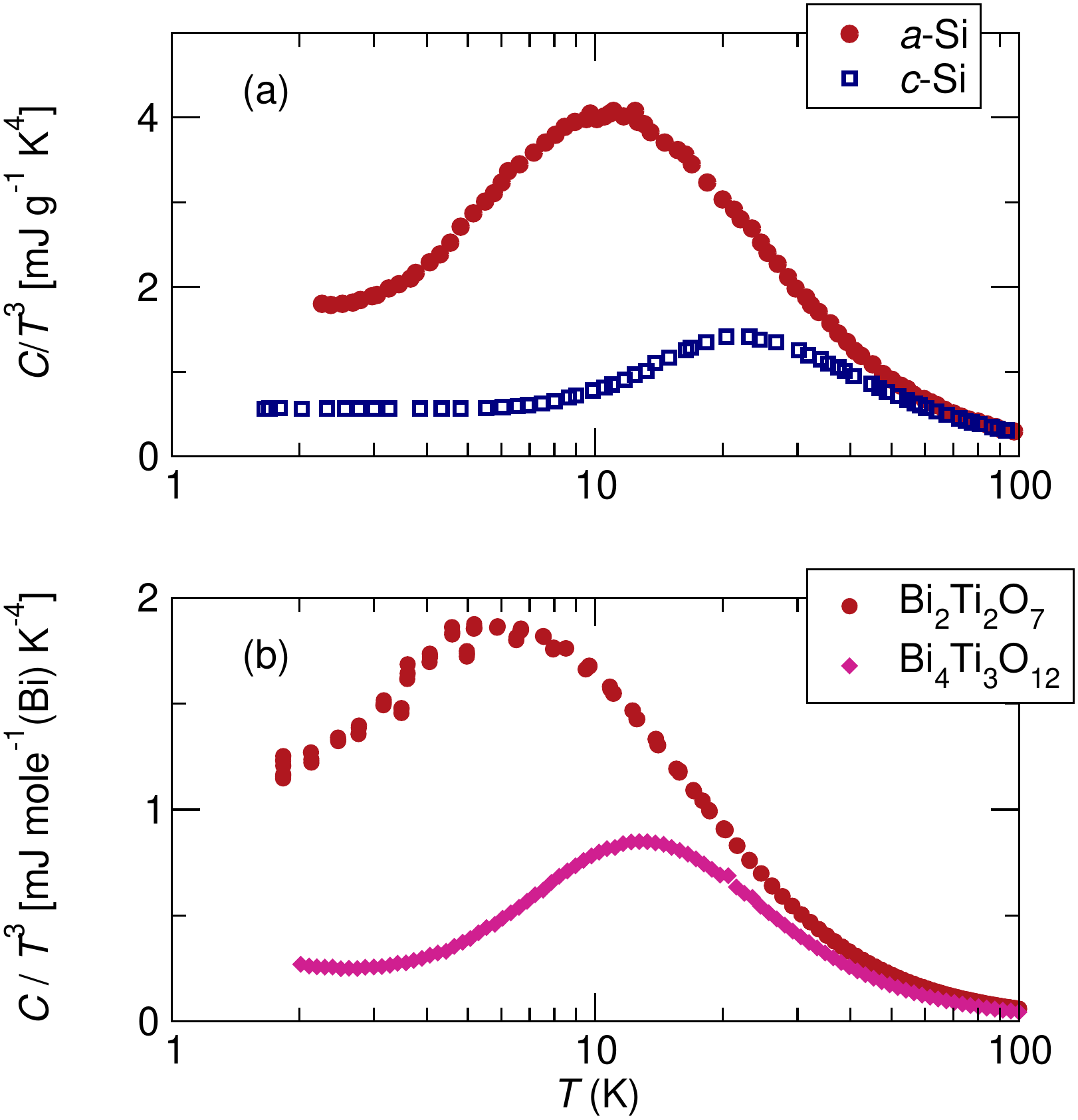}\\
\caption{(color on-line) (a) $C/T^3$ versus ln T amorphous SiO$_2$ and quartz\protect{\cite{pohl}}, (c) $C/T^3$ versus ln T for \yto\, (squares)
and \bto\, (stars).}
\label{fig:fig4}
\end{figure}

To more clearly demonstrate the influence of increasing disorder on the $C/T^3$ peak
in heat capacity in crystalline materials, we compare our measurements with some other 
low temperature heat capacity studies on ordered and disordered crystals.  We plot
the excess low temperature heat capacity as $C/T^3$ versus ln T for amorphous SiO$_2$
and crystalline quartz in FIG.\,\ref{fig:fig4}(a), taken from Ref. \cite{pohl}, and for \btofe\, and \bto\, in
 FIG.\,\ref{fig:fig4}(b).  The lattice disorder present in  amorphous SiO$_2$ leads to a 
 reduction in temperature of the excess $C/T^3$ heat capacity peak as compared to structurally
 well-ordered quartz; this peak shift is a signature of increasing disorder in otherwise
 similar systems.
     Qualitatively, this is similar to the shift in peak temperature between
  \btofe\, and \bto, despite the fact that both of these materials are crystalline and have very similar
  Debye temperatures and elemental constituents, and hence bare oscillator spectral weights, that are roughly the same.
   These measurements provide thermodynamic justification to support  the suggestion that crystalline \bto\, 
 has additional disorder as compared to \btofe.  
 
 Because there is no evidence for any lattice disorder in \bto, the presence of a 
 low-temperature $C/T^3$ peak provides
 empirical evidence suggesting that the lone pair electrons on Bi are disordered due to geometrical frustration, consistent
 with previous XRD studies\cite{Hector_JSSC04}. 
 Increasing disorder in other crystalline systems
has been observed to shift this heat capacity peak to lower temperatures, including
 oxides with chemical site disorder, 
such as Ca replacing Sr in the CaSrFeCoO$_5$ brownmillerite\cite{brownmillerite}. 
This disorder-induced shift
 is also observed in metallic systems, although
the size of the effect may be significantlly smaller than in insulators\cite{Safarik_PRL06}.  
It has also been noted that geometrical frustration present in systems having
underconstrained degrees of freedom, such as ZrW$_2$O$_8$, can also lead to excess
low temperature heat capacity in crystalline systems\cite{Ramirez_PRL98}.
 These results suggest that the downshift of the specific heat peak in 
displacively disordered, but still crystalline, \bto\, may be associated with
the development of a frozen ``charge-ice'' state in these systems driven by geometrical
frustration, rather than the frozen glassy state associated with structural disorder.
Furthermore, by investigating the detailed phonon density of states in \bto\, and isostructural
\yto\, and correlating these results with heat capacity data, it may be possible to gain insight
into the microscopic mechanisms giving rise to excess low temperature entropy in disordered systems.

In summary, we find that several insulating crystalline materials with 
lone pairs, and with the pyrochlore, perovskite, and Aurivillius crystal
structures, exhibit an excess low temperature specific heat above the Debye 
background. The specific heat peaks 
in pyrochlores \bto\, and \bnio\, are however,
distinctly broader and stronger, and are found to appear at lower temperatures 
than those of isostructural \yto\, and compounds showing coherent 
coherent lone pair displacements (\pto\, and \btofe).  We attribute this suppression
of the excess specific heat to the presence of additional disorder in \bto\, and \bnio\,
arising from incoherent frozen displacements of the Bi lone pair electrons.  This suggests
that \bto\, may be an example of a structurally well-ordered system with charge disorder
introduced by geometrical frustration on the pyrochlore lattice.     

This work was supported by the National Science Foundation under NSF CAREER
DMR-06044823 (to GL) and DMR-0449354 (to RS), and by the Institute for
Materials Research at Wayne State University. Work at Santa Barbara made use of
MRSEC facilities supported by the NSF through DMR-0520415.

\end{document}